\documentclass{iau}
\usepackage{graphicx}

\title[
Four New Self-lensing Binaries from {\it Kepler}] 
{Four New Self-lensing Binaries from {\it Kepler}: Radial Velocity Characterization and Astrophysical Implications}

\author[Masuda et al.]   
{Kento Masuda,$^1$
Hajime Kawahara,$^{2,3}$
David W. Latham,$^4$ Allyson Bieryla,$^4$
Morgan MacLeod,$^4$
Masanobu Kunitomo,$^5$ 
Othman Benomar,$^6$ 
Wako Aoki$^6$ 
}

\affiliation{$^1$Institute for Advanced Study, Princeton, NJ 08540, USA; {\tt kmasuda@ias.edu} 
\\[\affilskip]
$^2$Department of Earth and Planetary Science, The University of Tokyo, Tokyo 113-0033, Japan
\\[\affilskip]
$^3$Research Center for the Early Universe, School of Science, The University of Tokyo, Tokyo 113-0033, Japan; {\tt kawahara@eps.s.u-tokyo.ac.jp},
\\[\affilskip]
$^4$Center for Astrophysics $|$\,Harvard \& Smithsonian, Cambridge, MA 02138, USA
\\[\affilskip]
$^5$Department of Physics, School of Medicine, Kurume University, Fukuoka 830-0011, Japan
\\[\affilskip]
$^6$National Astronomical Observatory of Japan, 2-21-1 Osawa, Mitaka, Tokyo 181-8588, Japan
\\[\affilskip]
}

\pubyear{2019}
\volume{357}  
\jname{White Dwarfs as probes of fundamental physics and tracers of planetary, stellar \& galactic evolution}
\editors{}
\begin{document}

\maketitle

\begin{abstract}

In \cite{2018AJ....155..144K} and \cite{2019ApJ...881L...3M}, we reported the discovery of four self-lensing binaries consisting of F/G-type stars and (most likely) white dwarfs whose masses range from 0.2 to 0.6 solar masses. Here we present their updated system parameters based on new radial velocity data from the Tillinghast Reflector Echelle Spectrograph at the Fred Lawrence Whipple Observatory, 
and the {\it Gaia} parallaxes and spectroscopic parameters of the primary stars.
We also briefly discuss the astrophysical implications of these findings.

\end{abstract}

\firstsection 
\section{Introduction}

Stars eclipsed by their compact object companions can display periodic brightening pulses in their light curves, rather than eclipse-like dips, due to gravitational microlensing. Such self-lensing binaries (SLBs) were discussed 50~years ago in the context of the search for ``collapsed stars" (e.g., \cite[{Trimble} \& {Thorne} 1969]{1969ApJ...156.1013T}), but had not been known until \cite{2014Sci...344..275K} reported the first such system, KOI-3278, from the {\it Kepler} photometric data. The system consists of a G-type star and a WD of $0.5\,M_\odot$ on a 88-day orbit.

The KOI-3278 system had been identified as a transiting planet candidate because the stellar light curves also showed repeating dips due to occultations of the WD, which triggered the planet-detection pipeline. On the other hand, the events with periods $\gtrsim1\,\mathrm{yr}$ were not the focus of this pipeline, although SLBs on wider orbits generally exhibit stronger signals because of the larger Einstein radius.
In \cite{2018AJ....155..144K} and \cite{2019ApJ...881L...3M}, we reported additional four SLBs with periods ranging from $419$ to $728\,\mathrm{days}$ from our search (\cite[{Kawahara} \& {Masuda} 2019]{2019AJ....157..218K}) for long-period transiting planets that show only one or two transits in the four-year photometric data from the prime {\it Kepler} mission. Here we report updated parameters of these SLBs based on the radial velocity (RV) data from the Tillinghast Reflector Echelle Spectrograph (TRES; \cite[{Szentgyorgyi} \& {Fur{\'e}sz} 2007]{2007RMxAC..28..129S}, \cite[{Mink} 2011]{2011ASPC..442..305M}) on the 1.5\,m telescope at the Fred Lawrence Whipple Observatory. We also briefly discuss the astrophysical implications of these systems.

\section{Updated System Parameters of the Self-lensing Binaries}

We identified KIC~3835482, KIC~6233093, KIC~12254688, and KIC~8145411, as new (i.e., second to fifth) self-lensing binaries in addition to the first system KOI-3278. The periods and phases of the orbits are well determined from self-lensing signals observed twice or three times in the {\it Kepler} data. The light curve of KIC 8145411 has a gap in the middle of the two detected pulses that originally caused a factor-of-two uncertainty in the period, but this degeneracy has been resolved with the RV data.

For KIC~3835482, KIC~6233093, and KIC~12254688, we derived masses and radii of the primary stars by fitting stellar models to their atmospheric parameters from TRES spectra (KIC~3835482) or LAMOST DR4 (the others; \cite[{Cui} {et~al.} 2012 and {Luo} {et~al.} 2015]{lamost1, lamost2}), 2MASS $K_s$-band magnitude, and the parallax from {\it Gaia} DR2, assuming $K_s$-band extinction from a dust map. The procedure is described in \cite{2019AJ....157..218K}. For KIC~8145411 without available {\it Gaia} parallax, we obtained a high-resolution spectrum using the 
High-Dispersion Spectrograph (\cite[{Noguchi} {et~al.} 2002]{2002PASJ...54..855N}) on the Subaru 8.2~m telescope
and derived the stellar parameters by comparing the spectrum to those of touchstone stars with well-determined physical properties (\cite[{Yee} {et~al.} 2017]{2017ApJ...836...77Y}).

\begin{figure}[htbp]
\begin{center}
 \includegraphics[width=0.495\textwidth]{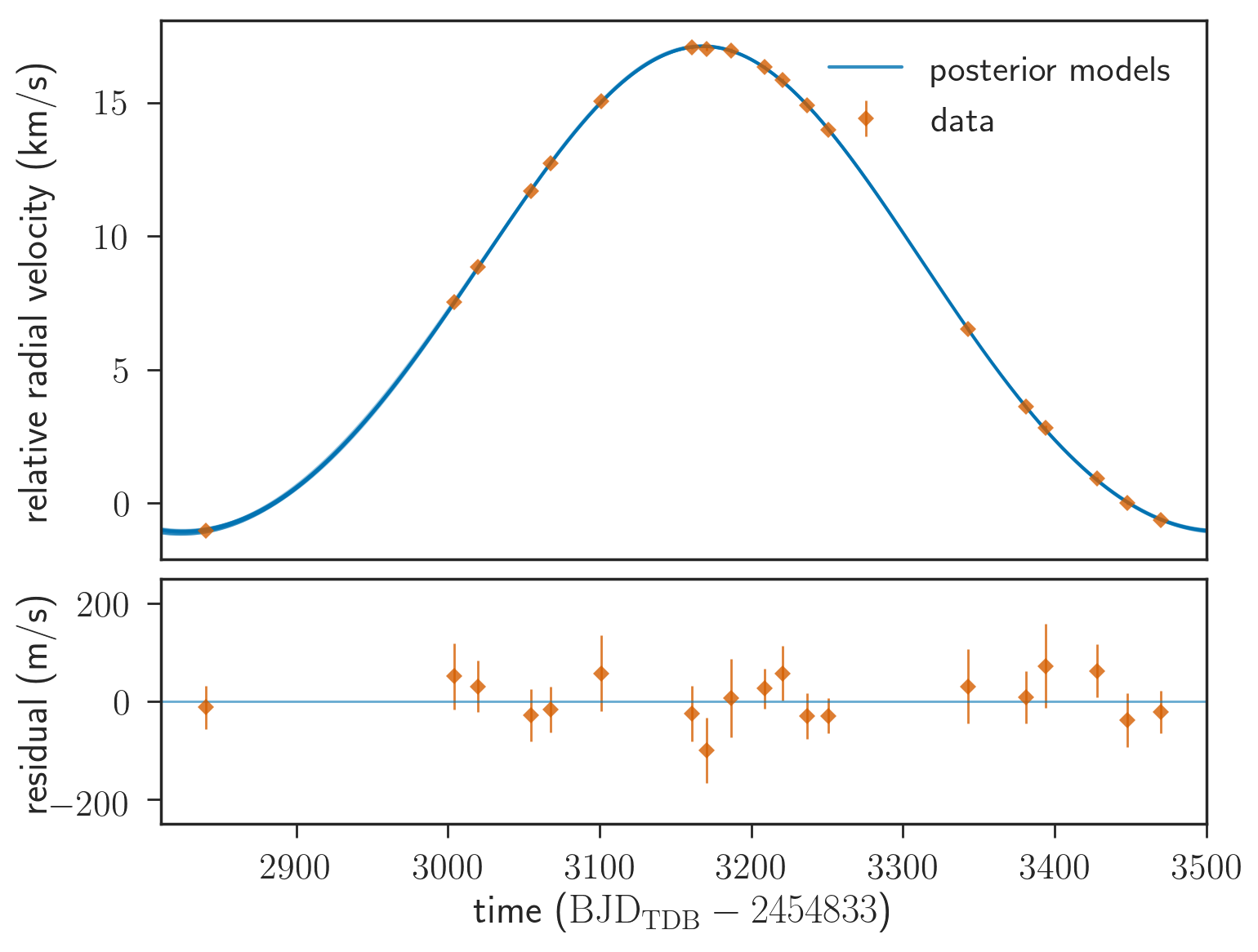} 
 \includegraphics[width=0.495\textwidth]{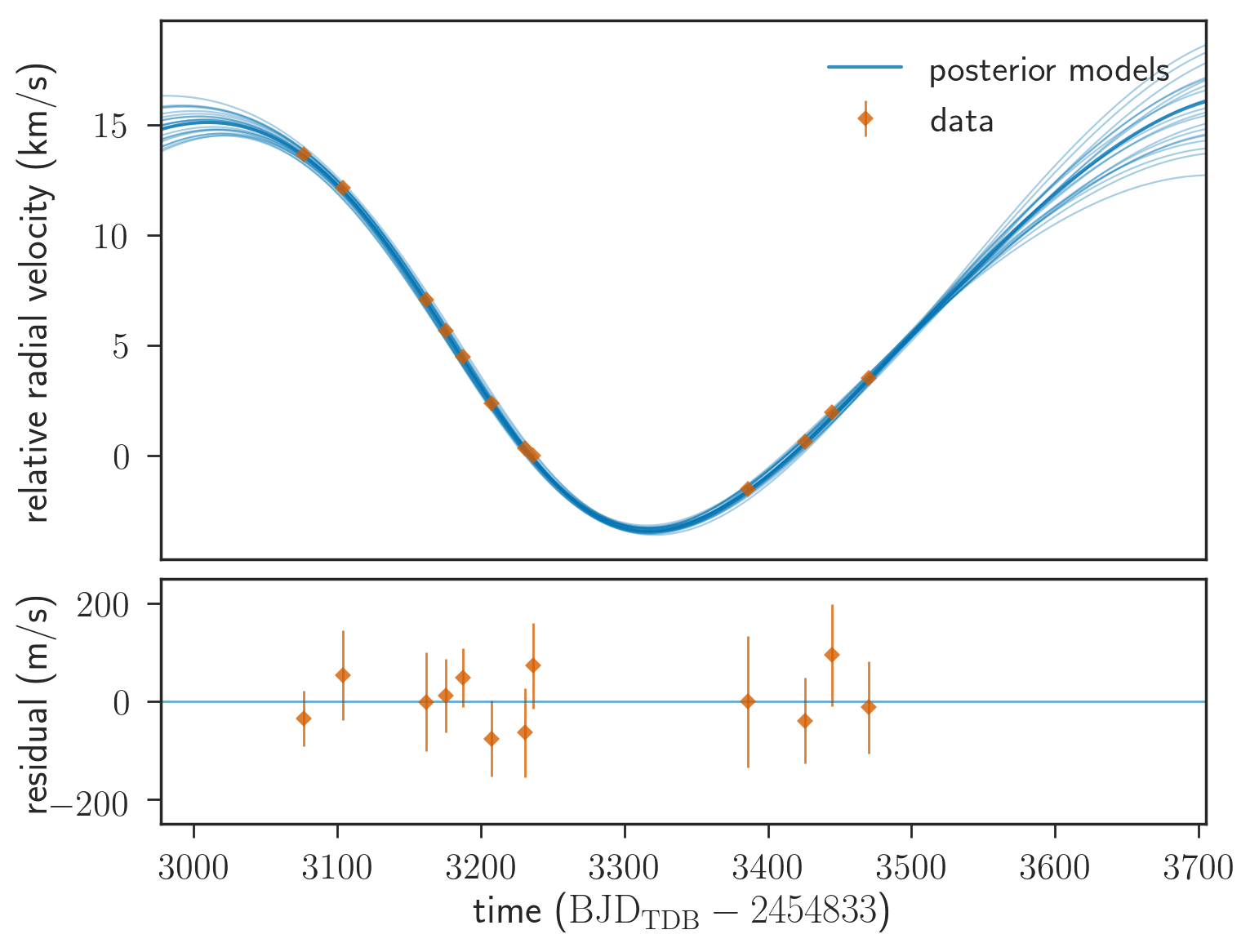} 
 \includegraphics[width=0.495\textwidth]{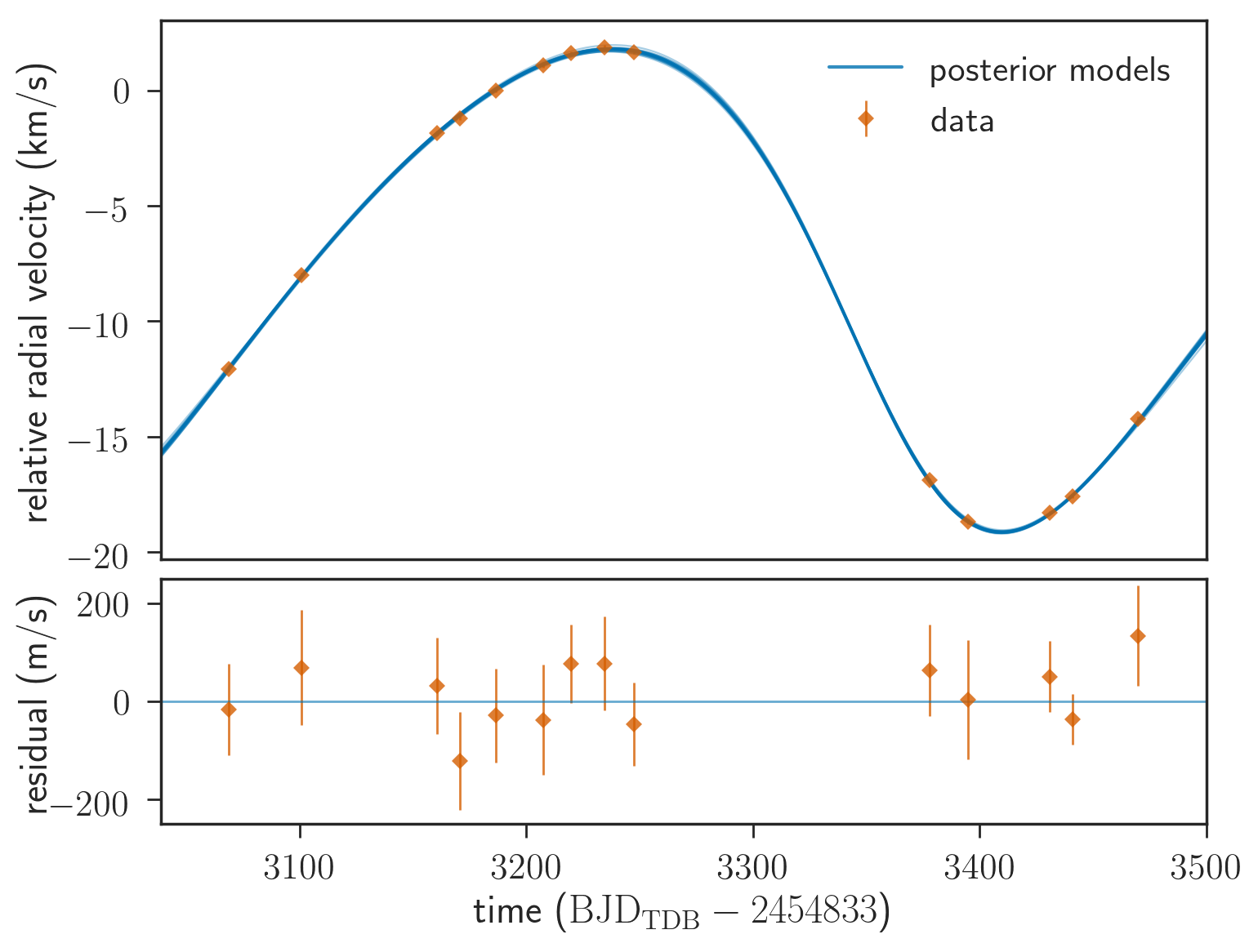} 
 \includegraphics[width=0.495\textwidth]{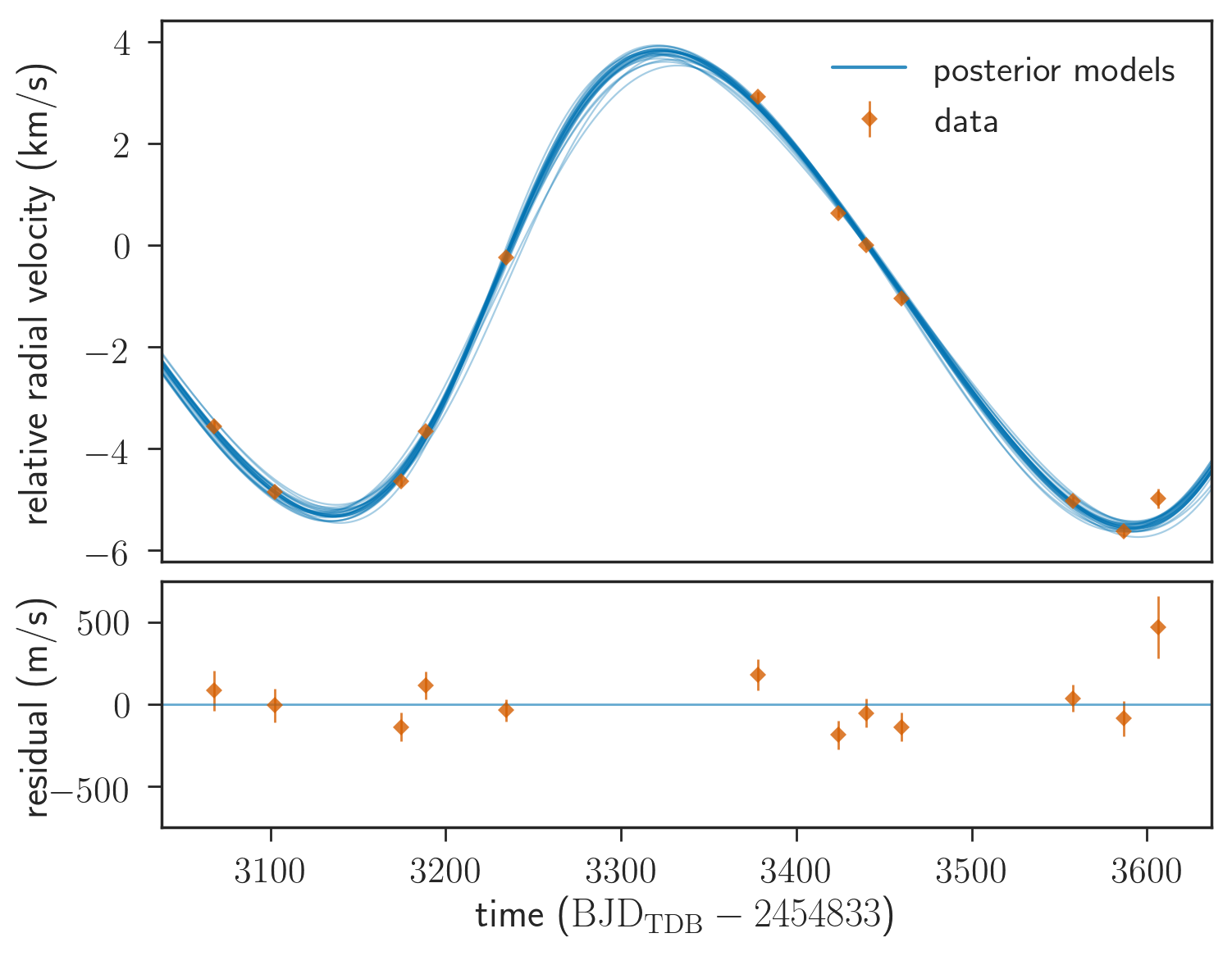} 
 \caption{Radial velocity data (orange diamonds) and posterior models (blue lines) for the four SLBs in \cite{2018AJ....155..144K} and \cite{2019ApJ...881L...3M}. {\it Top-left}---KIC~3835482.  {\it Top-right}---KIC~6233093. {\it Bottom-left}---KIC~12254688.  {\it Bottom-right}---KIC~8145411.}
   \label{fig2}
\end{center}
\end{figure}

We modeled TRES RVs of the SLBs with the prior information on their orbital periods and phases from the self-lensing light curves (Figure \ref{fig2}). The priors were otherwise chosen to be uninformative. The resulting binary mass functions were then combined with the primary masses estimated above to derive the WD masses assuming edge-on orbits (as justified from self-lensing). The results are summarized in Table \ref{tab:joint}.
 
\begin{table*}[h!]
\begin{center}
\caption{System Parameters of the Second to Fifth Self-lensing Binaries.\label{tab:joint}}
\begin{tabular}{lcccc}
\hline
& KIC~3835482 & KIC~6233093 & KIC~12254688 & KIC~8145411\\
\hline
\multicolumn{2}{l}{\it (Primary star)}\\
mass ($M_\odot$) & $1.52\pm0.06$ & $1.27\pm0.04$ & $1.37\pm0.07$ & $1.11\pm0.08$\\
radius ($R_\odot$) & $2.28\pm0.05$ & $2.32\pm0.07$ & $2.08\pm0.05$ & $1.27\pm0.18$\\
distance (kpc) & $1.15\pm0.02$ & $1.43\pm0.04$ & $1.20\pm0.03$ & $\cdots$ \vspace{0.2cm}\\  
\multicolumn{2}{l}{\it (Radial velocity)}\\
eclipse epoch & $591.628\pm0.005$ & $260.678^{+0.008}_{-0.009}$ & $404.583^{+0.006}_{-0.007}$ & $267.97\pm0.01$\\ 
\quad ($\mathrm{BJD_{TDB}}-2454833$)\\
orbital period $P$ (day) & $683.266^{+0.008}_{-0.006}$ & $727.979^{+0.010}_{-0.008}$ & $418.718^{+0.006}_{-0.005}$ & $455.83\pm0.02$\\
orbital eccentricity & $0.062^{+0.002}_{-0.001}$ & $0.123\pm0.008$ & $0.180^{+0.004}_{-0.003}$ & $0.13\pm0.02$\\
argument of periastron (deg) & $4\pm3$ & $112\pm10$ & $128\pm2$ & $-93\pm3$\\
RV semi-amplitude (km/s) & $9.07\pm0.02$ & $9.5\pm0.3$ & $10.44^{+0.06}_{-0.05}$& $4.59^{+0.09}_{-0.10}$\\
constant acceleration (m/s/day) & $0.1\pm0.1$ & $2^{+2}_{-3}$ & $-0.2^{+0.4}_{-0.3}$ & $-0.6\pm0.3$\\
RV zero point (km/s) & $7.45\pm0.01$ & $6.61^{+0.07}_{-0.06}$ & $-7.49^{+0.2}_{-0.3}$ & $-0.78^{+0.06}_{-0.07}$\\
RV jitter (m/s) & $13^{+12}_{-6}$ & $17^{+36}_{-9}$ & $17^{+29}_{-9}$ & $150^{+90}_{-80}$\vspace{0.2cm}\\
\multicolumn{2}{l}{\it (White dwarf)}\\
mass  ($M_\odot$) & $0.62\pm0.01$ & $0.61\pm0.02$ & $0.56\pm0.02$ & $0.19\pm0.01$\\
\hline
\end{tabular}
\end{center}
{Note --- Quoted values are the median and symmetric $68.3\%$ interval of the marginal posteriors.}
\vspace{-1cm}
\end{table*}

\section{Discussion}

\begin{figure}[htbp]
\vspace*{-0.1cm}
\begin{center}
 \includegraphics[width=\textwidth]{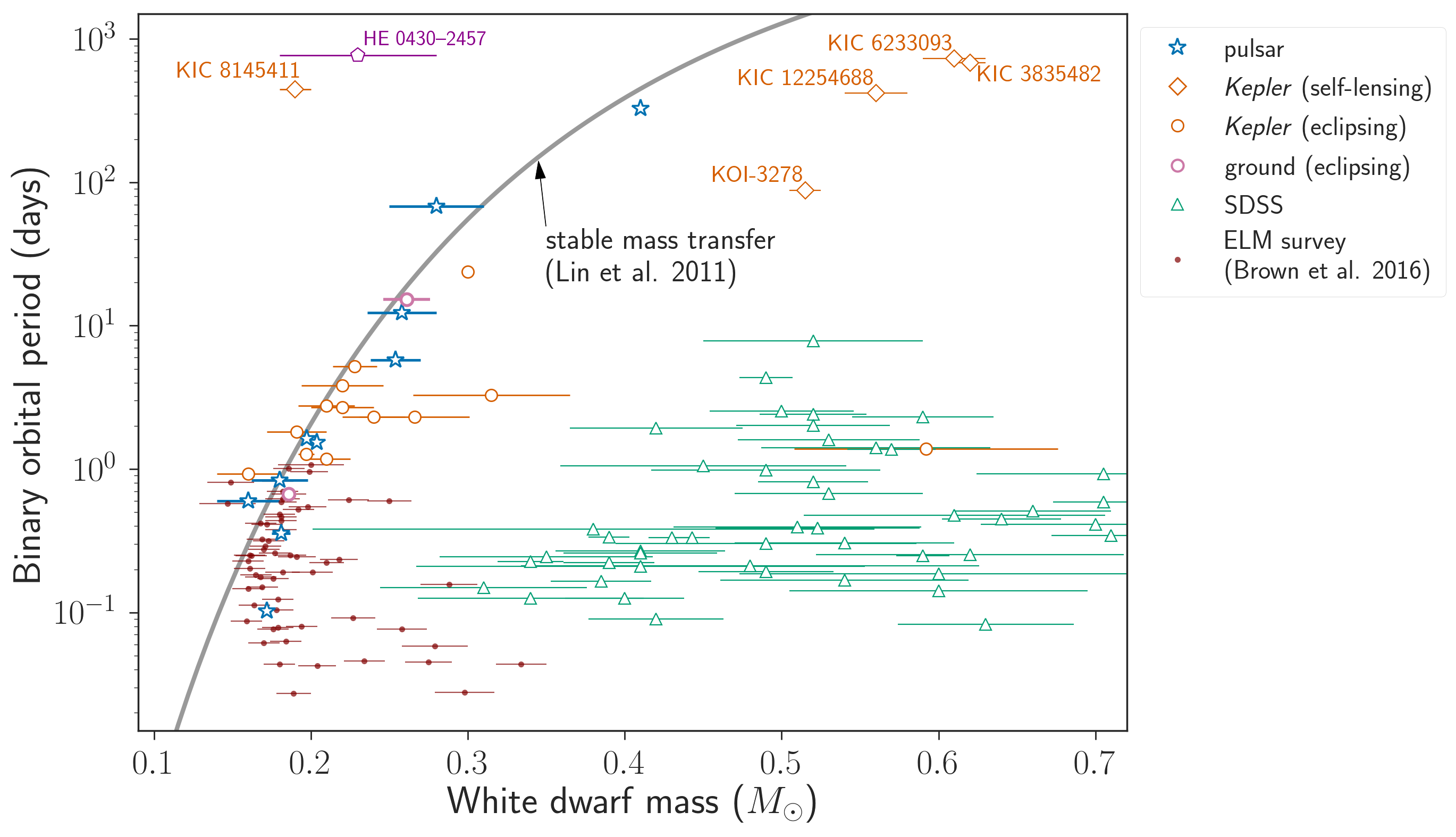}
 \caption{The masses of WDs in binaries and their orbital periods. Same as Figure~5 of \cite{2019ApJ...881L...3M}, but with updated parameters for SLBs (orange diamonds) from this work and \cite{2019ApJ...880...33Y}. Here we also show the ELM ``pre-WD" in the HE~0430--2457 system (\cite[Vos et al. 2018]{2018MNRAS.477L..40V}) with a purple pentagon.}
   \label{fig3}
   \vspace*{-0.7cm}
\end{center}
\end{figure}

Figure \ref{fig3} compares the binary orbital periods and WD masses of the five known SLBs against other known WD (or sub-dwarf star) binaries with well-measured masses; see Figure~5 of \cite{2019ApJ...881L...3M} for more details. The solid gray line shows the theoretical period--WD mass relation (\cite[Lin et al. 2011]{2011ApJ...732...70L}) for the case where the WD progenitor underwent stable mass transfer onto the companion. The good agreement between this prediction and the upper envelope of the measurements for WDs with $\lesssim0.3\,M_\odot$ supports the idea that these extremely low-mass (ELM) WDs are the products of binary interactions (\cite[Marsh et al. 1995]{1995MNRAS.275..828M}): such a low-mass WD cannot be formed in isolation because its progenitor would live too long to complete its evolution within the age of the Galaxy. 

The ELM WD with $0.2\,M_\odot$ in the KIC~8145411 system challenges this view. The orbit of this system is more than 10 times wider than theoretically expected, which means that the progenitor of this WD, when its core mass was $0.2\,M_\odot$, should have been far too small to have filled its Roche lobe. We do not have any good explanation of how this ELM WD could have been formed --- although the low orbital eccentricity of this system may still suggest that the binary interacted at some point in the past. Moreover, such systems may not be very rare among WD--stellar binaries on au-scale orbits. The system was found because the edge-on geometry allowed us to see self-lensing, and this happens only in 1 in 200 systems for the measured orbital separation. This implies that there are some 200 such ELM WDs around $10^5$ {\it Kepler} target stars, and this rate could be some 10\% of all WDs (i.e., regardless of their masses) on au-scale orbits around Sun-like stars. Indeed, a similar system containing an ELM ``pre-WD" has recently been reported by \cite{2018MNRAS.477L..40V} (purple pentagon in Figure \ref{fig3}), and some of the field blue straggler binaries might also contain such low-mass WDs. See Section~5 of \cite{2019ApJ...881L...3M} for more detailed discussion on the occurrence rate of such systems.

The other four SLBs (KOI-3278, KIC~3835482, KIC~6233093, KIC~12254688) have WDs with rather normal masses (0.5--$0.6\,M_\odot$), but their locations in Figure \ref{fig3} (i.e., they are below the gray line) suggest that these binaries should have interacted when the WD progenitors were on the asymptotic giant branch. Their low orbital eccentricities support this notion, although the latter three systems do have small but non-zero eccentricities and their origin is unclear. In this regard, these systems are reminiscent of some of the blue straggler binaries in open clusters and in the field for which stable mass transfer has been proposed to explain their bluer colors than expected for their ages. In fact, the primary stars in the KIC~3835482, KIC~6233093, and KIC~12254688 systems are all more massive than typical {\it Kepler} targets, although their ages are not well constrained.
Self-lensing enables detailed characterization of such post-interaction binaries by directly showing that the companion is a WD and by providing its precise true mass.

\end{document}